\documentstyle[12pt]{article}

\begin{document}
\sloppy
\begin{flushright}{SIT-HEP/TM-2}
\end{flushright}
\vskip 1.5 truecm
\centerline{\large{\bf Electroweak baryogenesis and hierarchy}}
\vskip .75 truecm
\centerline{\bf Tomohiro Matsuda
\footnote{matsuda@sit.ac.jp}}
\vskip .4 truecm
\centerline {\it Laboratory of Physics, Saitama Institute of
 Technology,}
\centerline {\it Fusaiji, Okabe-machi, Saitama 369-0293, 
Japan}
\vskip 1. truecm
\makeatletter
\@addtoreset{equation}{section}
\def\theequation{\thesection.\arabic{equation}}
\makeatother
\vskip 1. truecm
\begin{abstract}
\hspace*{\parindent}
We consider  the scenario of electroweak baryogenesis mediated by 
cosmological defects in a model of extra dimension.
We consider the domain wall on the brane in higher-dimensional
theories.
The electroweak breaking scale is suppressed
 and the sphaleron interaction is activated in the false vacuum.
\end{abstract}

\newpage
\section{Introduction}
\hspace*{\parindent}
Contrary to a naive cosmological expectation, all evidences suggest
that the Universe contains an abundance of matter over
antimatter.
Electroweak baryogenesis is an attractive idea in which
testable physics, present in the standard model of
electroweak interactions and its modest extensions,
is responsible for this fundamental cosmological
datum.
One may take the previous negative results as indication
that the asymmetry in the baryon number was not created
at the electroweak epoch, but rather related to the
physics of $B-L$ violation and neutrino masses.
To stick to electroweak baryogenesis one can consider
extensions of the particle content of the model
to get stronger electroweak phase transition.
In general scenario for electroweak baryogenesis requires
the co-existence of regions of large and small
$<H/T>$, where $H$ denotes the Higgs field in the 
standard model.
At small $<H/T>$, sphalerons are unsuppressed
and mediate baryon number violation while large
$<H/T>$ is needed to store the created baryon
number.
Below the critical temperature $T_{c}^{EW}$ of the electroweak
phase transition, $<H/T>$ grows until sphalerons are
shut-off.
For electroweak baryogenesis to be possible, one needs some
specific regions where $<H>$ is displaced from the
equilibrium value.

The idea we point out in this letter is that this can happen along topological
defects left over from some other cosmological phase
transitions that took place before the electroweak
phase transition\cite{trodden}.
If the electroweak symmetry breaking scale is suppressed  in some regions
around cosmological defects, sphalerons could be activated in such regions
 while they would be suppressed in the other part of space.
The motion of the defect network, in a similar way as
the motion of string surface in the usual defect-mediate
 scenario\cite{trodden}, will leave a net baryon number
behind the moving surface and then the baryon asymmetry will be kept 
in the sphaleron-suppressed true vacuum.
We find that a new type of defect-mediated electroweak baryogenesis
is possible when the radion is stabilized by the Goldberger-Wise 
mechanism\cite{GW}.
The mechanism for baryon number production at the phase boundary
is the same as the conventional defect-mediated electroweak baryogenesis.
The electroweak phase transition itself
is not required to be first order, which is the same characteristic of
 the conventional defect-mediated electroweak baryogenesis.
The strings-mediated electroweak baryogenesis is critically analyzed
in ref.\cite{string_EB} and the baryon number production by strings is proved
to be too small.
The critical point in ref.\cite{string_EB} 
is that the baryon number violation in 
the string core is too slow since the region
where electroweak symmetry is restored is never wide enough to
allow sufficient sphaleron events.
Such suppression does not appear in our case, which makes it possible
to expect the generation of the observed baryon number by
defect-mediated electroweak baryogenesis.

\section{Defect-mediated electroweak baryogenesis in GW model}
\hspace*{\parindent}
When the hierarchy is determined by the typical length
scales of the extra dimensions, there must be some 
mechanisms that ensure the stability of such scales.
If the mechanism for the stability is affected by the
defects on the brane or in the bulk, the defects may induce the
displacement of the electroweak scale in the defect core or in the
false vacuum,
resulting in the same mechanism discussed in ref.\cite{trodden}.

Here we examine an attractive model proposed by 
Goldberger and Wise\cite{GW} for giving the radion a potential
energy to stabilize the length scale.
They introduced a bulk scalar field with different VEV's,
$v_{0}$ and $v_{1}$, on two branes.
If the mass $m$ of the scalar is small compared to the scale
$k$ which appears in the warp factor $e^{-ky}$, then it is
possible to obtain the desired interbrane separation and
one finds the relation $e^{-ky}\simeq (v_{1}/v_{0})^{4k^{2}/m^{2}}$. 
They added to the model a scalar field $\Phi$ with the following
bulk action
\begin{equation}
S_b={1\over 2}\int d^4 x\int_{-\pi}^\pi d\phi \sqrt{G} 
\left(G^{AB}\partial_A \Phi \partial_B \Phi - m^2 \Phi^2\right),
\end{equation}
where $G_{AB}$ with $A,B=\mu,\phi$ is given by\cite{RS}
\begin{equation}
 d^{2}s=e^{2kr_{c}|\phi|}\eta_{\mu\nu}dx^{\mu}dx^{\nu}-r_{c}^{2}
d\phi^{2}.
\end{equation}
They also included interaction terms on the hidden and visible branes
(at $\phi=0$ and $\phi=\pi$ respectively) given by
\begin{equation}
\label{23}
S_h = -\int d^4 x \sqrt{-g_h}\lambda_h \left(\Phi^2 - v_h^2\right)^2,
\end{equation}
and
\begin{equation}
\label{24}
S_v = -\int d^4 x \sqrt{-g_v}\lambda_v \left(\Phi^2 - v_v^2\right)^2,
\end{equation}
where $g_h$ and $g_v$ are the determinants of the induced metric on the 
hidden and visible branes respectively.  
The terms on the branes cause $\Phi$ to develop a $\phi$-dependent
vacuum expectation value $\Phi(\phi)$ which is determined classically 
by solving the differential equation
\begin{eqnarray}
0 &=& -{1\over r_c^2}\partial_\phi\left(e^{-4\sigma}\partial_\phi\Phi\right)
+m^2 e^{-4\sigma}\Phi + 4e^{-4\sigma}\lambda_v\Phi 
\left(\Phi^2 - v_v^2\right)\frac{\delta(\phi-\pi)}{r_c}\nonumber \\
& & \mbox{} + 4e^{-4\sigma}\lambda_h\Phi \left(\Phi^2 - v_h^2\right)
\frac{\delta(\phi)}{r_c},
\end{eqnarray}
where $\sigma(\phi)=kr_{c} |\phi|.$  
Away from the boundaries at $\phi=0,\pi$, this equation has the 
general solution
\begin{equation}
\Phi(\phi) = e^{2\sigma}[A e^{\nu\sigma}+B e^{-\nu\sigma}],
\end{equation}
with $\nu=\sqrt{4+m^2/k^2}$.  
Putting this solution back into the scalar field action and 
integrating over $\phi$ yields an effective four-dimensional 
potential for $r_c$.
Then the unknown coefficients $A$ and $B$ are determined by 
imposing appropriate boundary conditions on the 3-branes.
They considered the simplified case in which the parameters 
$\lambda_h$ and $\lambda_v$ are large, and supposing that $m/k\ll 1$, 
and neglecting the subleading powers of $\exp (-kr_{c}\phi)$, then
obtained the potential as
\begin{equation}
\label{pot}
V_\Phi(r_c)= k\epsilon v_h^2 + 4ke^{-4kr_{c}\pi}(v_v - v_h 
e^{-\epsilon kr_{c}\pi})^2\left(1+\frac{\epsilon}{4}\right) - 
k\epsilon v_h e^{-(4+\epsilon)kr_{c}\pi}(2 v_v - v_h e^{-\epsilon 
kr_{c}\pi})
\end{equation}
where terms of order $\epsilon^2$ are neglected.
If one ignores the terms proportional to $\epsilon$, 
this potential has a minimum at 
\begin{equation}
k r_{c} = \left(\frac{4}{\pi}\right) \frac{k^2}{m^2} 
\ln\left[\frac{v_h}{v_v}\right].
\end{equation}
With $\ln (v_h/v_v)$ of order unity, one only needs $m^2/k^2$ of 
order $1/10$ to get $kr_{c}\sim 10.$  

In this limit, it is energetically favorable 
to have $\Phi(0)=v_h$ and $\Phi(\pi)=v_v$. 
The configuration that has both VEVs of the same sign has lower 
energy than the one with alternating signs, and therefore corresponds 
to the ground state.  

Then a question arises:{\it 
``What happens if the vacuum with alternating signs is also produced at 
an early stage of the Universe?''}
Then from eq.(\ref{pot}), one can easily find that each term in the
effective potential has
the same positive sign and it looks like a runaway potential for
such an unstable configuration.
From eq.(\ref{23}) and eq.(\ref{24}), one may think that there are
discrete symmetries $Z_{2}^{h}\times Z_{2}^{v}$.
($Z_{2}^{h}$:$\Phi(0)=v_{h}\leftrightarrow \Phi(0)=-v_{h}(Z_{2}^{h})$ and  
$Z_{2}^{v}$:$\Phi(\pi)=v_{v}\leftrightarrow \Phi(\pi)=-v_{v}(Z_{2}^{v})$.)
This symmetry is, however, explicitly broken by the interaction terms
and only a discrete symmetry $Z_{2}'$, which corresponds to the
simultaneous flip is left.
(See also eq.(\ref{pot}) and ref.\cite{GW}.)
In the conventional case, the potential in the hidden brane is so high that 
one can assume $\Phi(0)=v_{h}$.
In this case the domain wall is induced by $v_{v}$.
The explicit breaking term is explicitly given in eq.(\ref{pot}).
Substituting $v_{v}$ by $-v_{v}$, one can find the false vacuum
potential for $r_{c}$.
In general, an instability of the vacuum is a problem if the vacuum
is the true vacuum or it dominates the whole universe.
However, in our paper the unstable vacuum is a 
false vacuum which appears in the universe as the bubble surrounded by
the true vacuum.
In this case, the unstable false vacuua disappears before they become
 harmful.
What we should be concerned about is the local behaviour of the effective
theory around the phase boundary.
In the true vacuum the positioning of the brane is not altered.
In the false vacuum, the positioning of the brane is altered to make the
brane distance larger than the true vacuum.
At the temperature near the electroweak phase transition, the symmetry
restoration can be induced by the small shift of the effective electroweak 
scale.
In this respect, the phase boundary becomes much thinner than the
background wall configuration because of its
exponential dependence on the brane distance.
What we should be concerned about is the higgs profile which induces
the flux in front of the phase boundary, since it controls the
electroweak baryogenesis.
The domain wall that interpolates the vacuum with $\Phi(\pi)=v_v$ and
$\Phi(\pi)=-v_v$ at the visible brane 
(or possibly $\Phi(0)=v_h$ and $\Phi(0)=-v_h$ at the hidden brane)
is nothing but the commonly known  $Z_{2}$ domain wall with explicit 
breaking of $Z_{2}$ symmetry in the effective four dimensional theory.

For electroweak baryogenesis to be possible, as we have noted, one needs some
specific region where Higgs vacuum expectation value
 $<H>$ is displaced from the equilibrium value.
Here the difference of the warp factor is expected to induce the
difference of the electroweak scale in the local region.
Now we consider the case where the electroweak symmetry breaking scale 
is suppressed in the false vacuum. 
Then sphaleron interactions are activated in this
restricted area  while they are suppressed in the bulk of space 
when $T<T_{EW}$.
The motion of the defect network, in a similar way as
the motion of bubble walls in the usual strongly first order
phase transition scenario, will leave a net baryon number
behind the moving surface and then the baryon asymmetry will be kept 
in the sphaleron-suppressed regions.
Although the defect should have a long tail toward the runaway
direction, the typical length scale that is relevant for the
electroweak baryogenesis is determined by 
the radion mass at the phase boundary, which is larger than
the Higgs mass.
In this sense, the changes in the effective electroweak scale induced
by the background defect configuration is steep so that
the thin wall approximation is possible when one considers
 the electroweak baryogenesis.
The mechanism for baryon number production is the same as the conventional
defect-mediated electroweak baryogenesis.
The electroweak phase transition itself
is not required to be first order, which is the same characteristic of
 the conventional defect-mediated electroweak baryogenesis.
Thus we conclude that the scenario is a possible candidate
for generating sufficient BAU.

Here we also mention the collapsing mechanism for the cosmological
domain wall.
When the collapse is induced by the energy difference $\epsilon$
which is induced by the explicit breaking of $Z_{2}$ symmetry,
one can add extra components in the bulk
to adjust $\epsilon$ to a suitable value.
Another mechanism is the biased domain wall\cite{bias}, whose decaying
 process  is determined by cosmology, and
may (or may not) be adjusted to produce the suitable
domain wall structure.

What we are considering is the electroweak baryogenesis,
in which the the flux is injected by the phase boundary into 
the unbroken phase and it induces 
the baryon asymmetry near the phase boundary in the unbroken phase.
Then the produced baryons are trapped in the broken phase.
In this respect, the mechanism of our model is similar to the
conventional mechanism for electroweak baryogenesis.
The efficiency of the mechanism is determined by the higgs profile
in the phase boundary.
In our case the phase boundary is thin and the magnitude of the
injected flux is determined by the CP phase and the vacuum expectation
value of the Higgs field in the broken phase\cite{add}.
The electroweak baryogenesis in the effective theory of the brane
universe does not utilize the bulk dynamics to produce the baryon
asymmetry, since the sphalerons are activated in the unbroken phase.

We should note that the radion stabilization is generally
affected by the potentials on the brane and in the 
bulk\cite{genst}.
In this respect, the defects in the bulk or on the brane
can act to displace the radion even if no specific mechanism is 
implicated, and there is a chance for our mechanism to work
in any models for radion stabilization.

\section{Conclusions and Discussions}
\hspace*{\parindent}
In this letter we have proposed a novel possibility for 
electroweak baryogenesis mediated by cosmological defects.

We considered an interesting aspect of the Goldberger-Wise
mechanism for the stabilization of the radion in the RS model.
We expect that this mechanism works in other models
for extra dimensions in which the radion is stabilized by
the configurations in the bulk or on the brane\cite{susyextra}.

The electroweak phase transition itself
is not required to be first order, which is the same characteristic of
 the conventional defect-mediated electroweak baryogenesis.
Although the problem of the first order phase transition
is solved in our model, the problem of small CP breaking
parameter remains.
To obtain large CP parameter, one should extend the
low energy effective theory.

\section{Acknowledgment}
We wish to thank K.Shima for encouragement,  and people in
Tokyo University for kind hospitality.

\newpage
\begin{figure}[h]
\caption{The cosmological domain wall that appears at the temperature
$T_{c}>T>T_{c}'$ is given.
Here the structure of interest is the domain wall on the visible brane,
which interpolates between $\Phi(\pi)=+v_{v}$ and $\Phi(\pi)=-v_{v}$.
The left hand side (true vacuum) is already in the broken phase,
but the right hand side (false vacuum) is still in the symmetric phase
because the effective scale of the electroweak symmetry breaking has a
 gap.
$H_{0}$ denotes the Higgs expectation value in the true vacuum.
Our mechanism works at the temperature $T_{c}>T>T_{c}'$,
where $T_{c}$ and $T_{c}'$ denote the critical temterature in the 
true vacuum and the one in the false vacuum.
The Higgs vacuum expectation value is $0$ in the false vacuum because
of the gap in the effective critical temperature which we have assumed to
be induced by the shift of the hierarchy factor.
Note that the walls does not necessarily sweep the whole Universe,
which is the similar situation as the conventional defect-mediated
 electroweak baryogenesis.}
\label{fig1}
\end{figure}

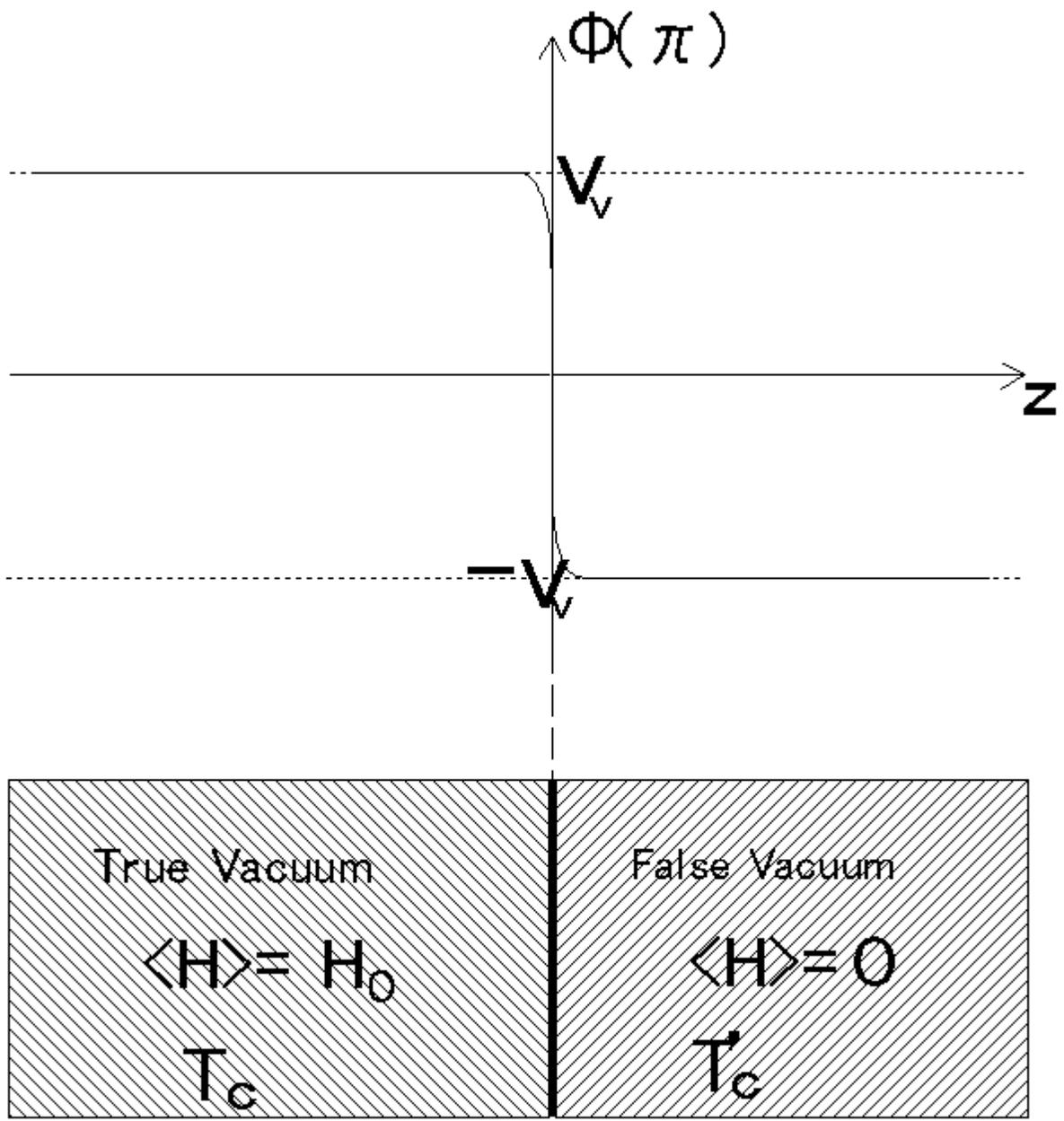
\end{document}